# Enhancement of long-wave vibronic interaction and quantum diffusion in liquids

V. Hizhnyakov, A. Shelkan

Institute of Physics, University of Tartu, W. Ostwald Street 1, 50411 Tartu
hizh@ut.ee

**Abstract**. The zero-phonon type hoping motion of defects in the solid and liquid phases, causing quantum diffusion, is considered. It was found that due to the previously discovered significant enhancement of vibronic interaction with long-wave acoustic phonons, this motion in liquids can be significantly amplified compared to the solid phase of the same substance at a close temperature. Quantum diffusion may be particularly important in superfluid $^4$He. This is especially true for vacancies that can move here in the form of vacancy waves (zero vacancies).

## 1. Introduction

It has recently been shown [1] (see also [2]) that due to the Archimedes principle, the linear vibronic interaction with long-wave acoustic phonons in liquids is greatly enhanced compared to the solid phase. Here we show that quadratic and higher-order vibronic interactions with these phonons are also amplified in liquids. The noted difference in the vibronic interaction in liquid and solid phases is especially significant for zero-phonon transitions. Indeed, in the solid phase, the probability of these transitions in the optical spectra of impurity centres in case of linear vibronic interaction is finite. However, in the liquid phase, this probability tends to zero [1,2]. As a result, instead of a discrete line in the optical spectra of impurity centres in liquids, including the superfluid component of the liquid helium , a broadened ZPL appears [1,2], which is consistent with experimental measurements [3] of the ZPL of the Dy atom in superfluid helium.

Zero-phonon transitions also contribute to the hopping movement of light atoms in media at low temperatures, causing sub-barrier tunnelling of particles. This process, called quantum diffusion (QD) [4-7], is described by the diffusion coefficient

$$D = a_0^2 \overline{\Delta}^2 / \gamma, \qquad (1)$$

where $\overline{\Delta} = \Delta e^{-S/2}$, $\Delta$ is hoping integral, $S$ is the Huang-Rhys factor describing the logarithm of the inverse probability of a zero-phonon hop, $\gamma$ is the spectral width of zero-phonon transitions, $a_0$ is the length of a hop. The factor $e^{-S}$ denotes the relative probability of zero-phonon transitions; the width $\gamma$ of the spectrum takes into account THE dynamic destruction of the energy band of particles due to distortion of the local dynamics of phonons during the hop. Equation (1) is valid in the typical case when the free path of a particle $l$ is less than $a_0$ [4-7], which happens if the time of hope $\overline{\Delta}^{-1}$ exceeds the time $\gamma^{-1}$ of destruction of coherent motion. In solid $^4$He (existing at pressure above 25 atmospheres) at very low temperatures, the opposite ratio is possible. Then QD occurs as the motion of delocalized quasiparticles [7]. QD of light atoms was observed in solid $^4$He at very low temperatures; see [8,9].

Note that the Huang-Rhys factor is mainly determined by the linear interaction with phonons; in contrast, the width of the spectrum $\gamma$ is determined solely by the quadratic interaction with phonons [10]. The latter interaction leads to the modulation of the energy difference of the initial and final levels by phonons, thereby destroying the energy band of the coherent motion of the particle [4,6,7]. The underlying quantum mechanism is the simultaneous creation and destruction of phonons during a jump called the Raman mechanism. This mechanism is especially important in cases when hops are accompanied by the rupture and restoration of strong bonds of the atom performing the hops (see [11,12], where this case of QD was considered).

Here we assume that sub-barrier tunnelling transitions can take place also in liquid phase, in particular, in superfluid helium, taking into account that the immediate environment of an impurity atom in this phase is largely the same as in the solid phase. As shown below, the probability of these transitions can greatly increase in the liquid phase compared to the solid phase of the same substance at a close temperature. This increase, in turn, can greatly enhance the QD of particles in the liquid phase. QD may be particularly important in superfluid He.

## 2. Model

We consider a single defect (impurity atom) that can hop from site to site in the atomic medium, both solid and liquid. To be specific, we consider the motion of the impurity atom along interstitial positions. The model we use to describe the motion of an impurity atom has been discussed in a number of papers considering QD in solids (see, e.g. [6,7,11,12]). In this model, as a basis one uses the localized states $|i\rangle$, where $i$ labels a lattice site, which can be occupied by the defect. In this representation the Hamiltonian of the system reads [11,12]

$$H = \sum_{(i,i')} \left( \Delta_{ii'}(1-\delta_{ii'}) + H_i \delta_{ii'} \right) |i\rangle\langle i'|, \qquad (2)$$

where $H_i$ is the Hamiltonian of phonons when the defect occupies the site i, $\Delta_{ii'}$ is the tunnelling Hamiltonian matrix element, which, in principle, may include phonon operators. We take into account the linear and quadratic interaction of the defect with the surrounding atoms (ions) taking $H_i$ in the form

$$H_i = H_{ph} + aQ_i + \frac{1}{2}bQ_i^2. \qquad (3)$$

Here $H_{ph}$ is the phonon Hamiltonian of the environment (described in harmonic approximation), $Q_i$ is the configurational coordinate of totally-symmetric displacements of the nearest neighbouring atoms from the impurity atom, $a$ is the parameter of the linear interaction, $b$ is the parameter of the quadratic interaction. The coordinate $Q$ can be expressed as $Q_i = \sum_j \tilde{e}_i x_j$, where $x_j$ are normal coordinates of phonons, $\tilde{e}_{i,j} = \sum_{i'_i} \left( e_{i'j} - e_{i,j} \right)$, $i'_i$ denotes the sites nearest to $i$, $e_{i,j}$ is the contribution of phonon number $j$ to the coordinate of an atom number $i$.

We consider an impurity atom surrounded by the nearest atoms in $\pm x, \pm y, \pm z$ directions at the distance $a/2$. Considering phonons as plain waves we obtain $\tilde{e}_{i,\vec{k}} = -i\left(2/\sqrt{3}\right)\sum_\alpha \sin(k_\alpha a/2)$, where $\alpha = x, y, z$, $k_\alpha$ is $\alpha$ component of the wave vector $\vec{k}$ of phonon with frequency $\omega = kva$, $v$ is velocity of sound (we use dimensional frequencies with $\omega = 1$ being the top phonon frequency). Using spherical coordinates for the wave vector $\vec{k}$ with $z$ axis in the direction of a nearest atom, we get the following equation for the dependence of $\left|\tilde{e}_{i,j}\right|^2$ on the frequency $\omega = \omega_j$:

$$\left|\tilde{e}_{1,\omega}\right|^2 = (2/\pi)e_\omega^2 \int_0^\pi d\varphi \sin^2(\omega\cos\varphi/2) = e_\omega^2\left(1 - J_0(\omega)\right). \qquad (4)$$

Here $J_0(\omega)$ is a zero-order Bessel function, $e_\omega^2 \equiv e_{1k}^2 \rho(\omega_k)$, $\omega_k = \omega$, $\rho(\omega)$ is the density of states of phonons; $e_\omega^2$ is normalized according to $\int_0^1 d\omega\, e_\omega^2 = 1$.

Taking into account only the term $H_i$ in the Hamiltonian (2) results in the following shifts of the equilibrium position of normal modes $j$: $u_{ij} = a\tilde{e}_{i,j}/\omega_j^2$. These shifts, in turn, lead to the displacement of all the atoms of the medium, including those located on its outer surface. In macroscopic crystals the latter shifts tend to zero so that the entire volume is not changed [1]. This is due to the fact that in crystals in the low frequency limit for both longitudinal and transverse phonons

$$e_{ij,Cr} \to Const \neq 0, \quad \tilde{e}_{ij,Cr} \propto \omega_j \to 0, \quad \omega_j \to 0 \qquad (5)$$

(the subscript $Cr$ means crystal) This gives $u_{ij,Cr} \propto \omega_j^{-1}$, $\omega_j \to 0$. For such $u_{ij,Cr}$, the volume of the macroscopic crystal described by the Hamiltonian $H_1$ does not change in comparison with the volume of a macroscopic crystal described by the Hamiltonian $H_{ph}$ [1].

Only acoustic phonons exist in liquids, which leads to the existence of the Archimedes principle. It was shown in [1] that due to this principle, the local force in the Hamiltonian $H_1$ leads to a finite (non-zero) change in the entire volume of the liquid, which gives $\omega_j^2 u_{ij,L} \to Const \neq 0$, $\omega_j \to 0$ and

$$e_{ij,L} \propto \omega_j^{-1}, \quad \tilde{e}_{ij,L} \to Const \neq 0, \quad \omega_j \to 0 \qquad (6)$$

(the subscript $L$ means liquid). Equations (5) and (6) express a significant difference in the vibronic interaction of crystal and liquid phases: the centre of gravity of this interaction with acoustic phonons in liquid phase is shifted to the long-wave side compared to crystal phase. Consequently, those phonon processes in which the contribution of low-frequency acoustic phonons is weakened decrease in the liquid phase. This difference in phonon processes in solid and liquid phases appears to be very important for the zero-phonon transitions: in the case of linear vibronic interaction, the probability of these transitions in the optical spectra is finite in solid phase, but it tends to zero in the liquid phase [1]. Here we show that, for this reason, the probability of zero-phonon transitions leading to sub-barrier tunnelling and QD in the liquid phase essentially increases compared to the crystalline phase.

### 3. Rate of tunnelling transitions

To calculate the rate of the tunnelling transition $\Gamma$, we use the Fermi golden rule. Then, taking into account only the independent on phonon coordinates part of the tunnelling matrix element, we get (see [6,11,12], where both phonon-independent and phonon-dependent parts of tunnelling matrix elements was taken into account)

$$\Gamma = 2\Delta^2 \operatorname{Re} \int_0^\infty F(t)dt; \quad F(t) = \left\langle e^{itH_2} e^{-itH_1} \right\rangle_1. \qquad (7)$$

Here $\Delta \equiv \Delta_{12}$ is the tunnelling matrix element, $H_1$ and $H_2$ are Hamiltonians of vibrations of atoms for impurity atom being in the first (before the hop) and in the second position (after the hop), respectively, $\langle...\rangle_1$ indicates the thermal averaging over eigenstates of Hamiltonian $H_1$,

$$V_1 = a(Q_1 - Q_2) = \sum_j \alpha_j x_j,$$

$$V_2 = \frac{1}{2}(Q_1 b Q_1 - Q_2 b Q_2) = \frac{1}{2}\sum_{j,j'} \beta_{jj'} x_j x_{j'},$$

(8)

$\alpha_j = a(\tilde{e}_{1j} - \tilde{e}_{2j})$, $\beta_{jj'} = b(\tilde{e}_{1j}\tilde{e}_{1j'} - \tilde{e}_{2j}\tilde{e}_{2j'})$, $\hat{T}$ is the time ordering operator, $x_j$ are normal coordinates of phonons, $\hbar = 1$. By applying the shift operators $e^{\pm\nabla}$ with $\nabla = \sum_j (u_{1j} - u_{2j})\partial/\partial x_j$ and $u_{ij} = \alpha_j/\omega_j^2$, the Hamiltonian $H_2$ gets the form [10,11]

$$H_2 = e^{-\nabla}\tilde{H}_2 e^{\nabla}, \quad \tilde{H}_2 = H_1 + V_2,$$

(9)

where $\tilde{H}_2$ differs from $H_1$ due to purely quadratic interaction.

At low temperatures the largest contribution to $\Gamma$ in Eq. (7) comes from the zero-phonon transitions determined by the asymptotic of $F(t)$ at large $t$. In this limit, in systems with a continuous phonon spectrum [11,12]

$$F(t) \cong \langle e^{\nabla}\rangle_1 \langle e^{-\nabla}\rangle_1 \langle e^{it\hat{H}_2} e^{iit H_1}\rangle_1.$$

(10)

Here $\langle e^{\nabla}\rangle_1 = \exp(-\langle\nabla^2\rangle_1/2)$. For a large time $t$, in the case of a purely quadratic interaction, $\tilde{F}(t) = \langle e^{it\tilde{H}_2} e^{-it H_1}\rangle_1$ exponentially decays with time [10, 13-16]. Therefore, in the large $t$ limit (see for example. the nonperturbative theory of zero-phonon transitions [10])

$$F(t) \cong \exp(-S - \gamma t),$$

(11)

where $S$ is the Huang-Rhys factor, describing contribution of the linear interaction to the probability of the zero-phonon hop transition, $\gamma$ gives the spectral width of zero-phonon transitions. In general case $S$ and $\gamma$ are complex functions of $a$, $b$ and the parameters of the phonon spectrum [10]. However, in the case of small $b$ these functions greatly simplify obtaining the form [10,13-15]

$$S = -\langle\nabla^2\rangle_1 = (1/2\hbar)\sum_j |\alpha_j|^2 \omega_j^{-3}(2\bar{n}_{\omega_j} + 1),$$

(12)

$$\gamma = (\pi/4\hbar^2)\sum_{jj'}\left(|\beta_{jj'}|^2/\omega_j\omega_{j'}\right)(n_{\omega_j}+1)n_{\omega_{j'}}\delta(\omega_j - \omega_{j'}),$$

(13)

where $n_\omega = 1/(e^{\omega/k_B T} - 1)$ is the Planck population factor of phonons. Note that the spectral width $\gamma$ arises as a result of participation in the transition of destruction and creation of phonons of the same frequency. This mechanism of broadening of the spectrum of zero-phonon transitions is called the Raman mechanism.

We are considering zero-phonon hops of an impurity atom in $z$-direction from the initial position $z = 0$ to the final position $z = a$ with the nearest surrounding atoms of cubic symmetry. Then in the low frequency limit

$$\alpha_j = a(\tilde{e}_{1j} - \tilde{e}_{2j}) = a\tilde{e}_{ij}\left(1 - e^{ik_{zj}a}\right) \propto a\tilde{e}_{ij}\omega_j, \quad \omega_j \to 0,$$

(14)

$$\beta_{jj'} = b(\tilde{e}_{1j}\tilde{e}_{1j'} - \tilde{e}_{2j}\tilde{e}_{2j'}) = b\tilde{e}_{ij}\tilde{e}_{1j'}\left(1 - e^{i(k_{zj}+k_{zj'})a}\right) \propto b e_{ij}\tilde{e}_{1j'}(\omega_j + \omega_{j'}), \quad \omega_j \to 0.$$

(15)

It follows from above equations that the contribution of long-wave acoustic phonons to $V_1$ and $V_2$ is reduced as compared to the one-site interactions $\alpha Q_{1,2}$ and $\beta Q_{1,2}^2$. This expresses an important property of phonon processes that cause transitions with hops, compared to those processes that occur without hops: long-wave phonons make a small contribution to phonon processes with hopes, while, unlike short-wave phonons, they do not distinguish well between close locations. It is important here that this effect of suppressing long-wave phonons for jumping is more important for the liquid phase than for the solid. This is due to the fact [1] that the contribution of low-frequency phonons to the vibrations of the atom is greater than in the solid phase.

To describe the effects of suppression of long-wave phonons during a hop in solid and liquid phases, we refine the frequency dependence of $\alpha_j^2 = a^2 |\tilde{e}_{1j} - \tilde{e}_{2j}|^2$ and $\beta_{jj'}^2 = b^2 |\tilde{e}_{1j}\tilde{e}_{1j'} - \tilde{e}_{2j}\tilde{e}_{2j'}|^2$. Using spherical coordinates for the wave vectors of phonons, we obtain $k_z = k\cos\varphi$ and

$$|\alpha_\omega|^2 = a^2 \left(\tilde{e}_\omega^2/\pi\right) \int_0^\pi d\varphi \left|1 - e^{i\omega\cos\varphi}\right|^2 = 2a^2 e_\omega^2 \left(1 - J_0(\omega)\right)^2, \qquad (16)$$

$$|\beta_{\omega\omega'}|^2 = \left(4\tilde{e}_\omega^2 \tilde{e}_{\omega'}^2/\pi^2\right) \int_0^\pi d\varphi \int_0^\pi d\varphi' \sin^2\left((\omega\cos\varphi + \omega'\cos\varphi')/2\right) \qquad (17)$$
$$= 2e_\omega^2 e_{\omega'}^2 \left(1 - J_0(\omega)\right)\left(1 - J_0(\omega')\right)\left(1 - J_0(\omega)J_0(\omega')\right).$$

Substituting these equations to equations (12) and (13) we obtain

$$S = a^2 \int_0^1 d\omega\, e_\omega^2 \omega^{-3} \left(1 - J_0(\omega)\right)^2 (2n_\omega + 1), \qquad (18)$$

$$\gamma = (\pi b^2/2) \int_0^1 d\omega\, e_\omega^4 \omega^{-2} (n_\omega + 1) n_\omega \left(1 - J_0(\omega)\right)^2 \left(1 - J_0^2(\omega)\right). \qquad (19)$$

To find these values, we use the Van Hove model of acoustic phonons. In this model

$$e_{\omega,Cr}^2 = (16/\pi)\omega^2 \sqrt{1-\omega^2},$$
$$e_{\omega,L}^2 = (4/\pi)\sqrt{1-\omega^2}. \qquad (20)$$

The difference in these expressions describes the shift of the "center of gravity" of the contribution of fully symmetrical acoustic phonons to the vibrations of atoms in the liquid phase towards low frequencies. The values of Huang-Rhys factor and the spectral width of the zero-phonon hop transitions in this model in their dependence on temperature are shown in Fig. 1

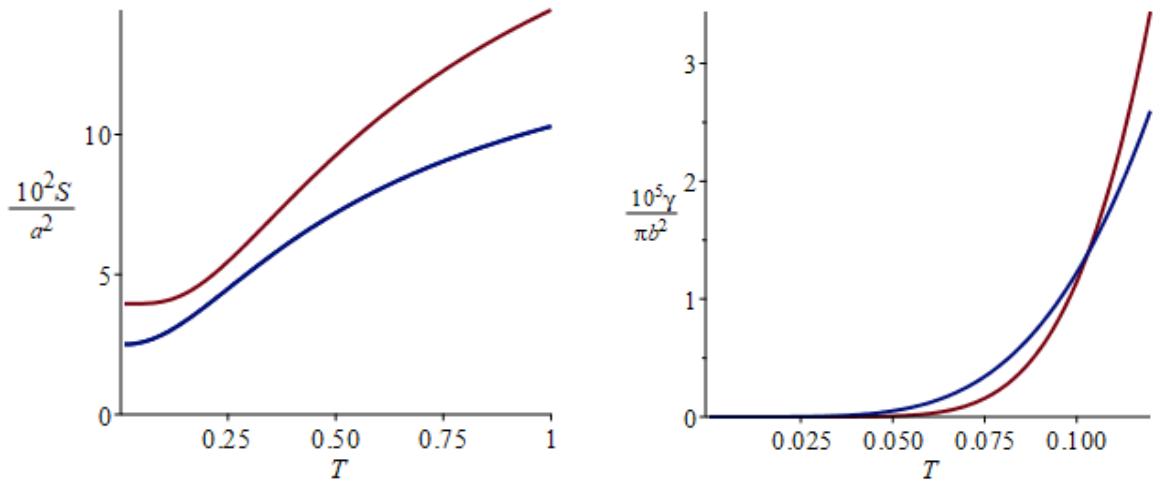

Fig. 1. Temperature dependence of the Huang-Rhys factor $S$ (left) and spectral width of the zero-phonon tunnelling transition (right) in the solid (broun lines) and in liquid phase (blue lines).

Note, that the used equation for $e_{\omega,L}^2$ does not correctly describe the disappearance of acoustic phonons at high frequencies in liquids. However, the contribution of high-frequency phonons to integrals (18) and (19) is small. Thus, the noted inaccuracy of the approximation used for the liquid phase does not significantly affect the calculation results.

## 4. Discussion

The results presented in Fig. 1 describe the temperature dependence – an increase in the Huang-Rhys factor and the width of the spectrum of zero-phonon transitions with an increasing in temperature. This increase leads to a rapid decrease in the diffusion coefficient of QD with increasing temperature. Because of this, QD can be significant only at low temperatures [4-7]. At the same time, in the liquid phase both $S$ and $\gamma$ increase more slowly with temperature than in the solid phase. In the low temperature limit, the dependence of QD on the temperature in the solid and liquid phases is completely determined by $\gamma$ at

$$\gamma_{Cr} \propto T^9, \quad \gamma_L \propto T^5, \quad T \to 0 \qquad (21)$$

(see [4], where the $T^9$-type temperature law for QD in crystals was first obtained).

As a rule, the Huang-Rhys factor largely exceeds one and can be several tens or more. For example, for the diffusion of $^3$He atoms in solid $^4$He at the observed value of $\bar{\Delta} \sim 10^6 \text{ s}^{-1}$ [4], $S/2 \sim 10$ can be estimated. In the case of heavier atoms and other liquids, $S$ can be much larger. At the same time, as follows from the results presented in Fig. 1 (left), the value of $S$ in the liquid phase is almost two times less than the value of $S$ in the solid phase of the same substance at a close temperature. This allows us to conclude that the probability $e^{-S}$ of the zero-phonon hops in liquid phase is many orders of magnitude larger than in solid phase of the same substance at close temperature. Liquid phase usually (except superfluid He) exists for temperatures above one tenth of maximum phonon frequency in solid phase. According to Fig. 1 (right), the value of $\gamma^{-1}$ in liquid phase also exceed the value of $\gamma^{-1}$ in solid phase. Besides, in liquid phase $e^{-S}$ and $\gamma^{-1}$ increase more slowly with temperature than in the solid phase. This also works in favor of liquid phase QD.

It should be noted that in addition to the Raman mechanism, there may be other mechanisms for broadening the spectral width of zero-phonon transitions. For example, the structure of the immediate environment of an impurity atom in a liquid may change over time. These fluctuations can also make an additional contribution to $\gamma$, if the rate of the change exceeds $\gamma$. There are also additional mechanisms of increasing $\gamma$ in the solid phase due to inhomogeneous static fields of various defects. Therefore the ratio of $\gamma$ values in solid and liquid phases depend on the used substance. However, a large difference of factors $\exp(-S)$ still works in favor of QD in the liquid phase.

A large increase in the factors $\exp(-S)$ in liquid phase as compared to the solid phase is especially important for the diffusion of defects in superfluid helium. Indeed, the bandwidth of the quantum particle in superfluid helium should be greatly increased compared to the solid He. In addition, in a superfluid He, the dynamic destruction of a particle band due to the destruction and creation of phonons during a quantum hop occurs only due to the presence of a normal component. The corresponding temperature dependence occurs extremely quickly because

$$\gamma \sim T^5 \rho_N \sim T^{11} \div T^{12}, \quad T \ll T_\lambda = 2.17 \, K \qquad (22)$$

(here $\rho_N \approx (T/T_\lambda)^{6.8}$ is the weight of the normal component [16]). Consequently, at very low temperatures, a light impurity atom in superfluid helium should behave like a delocalized quantum particle. This suggests that in superfluid helium, a Bose-Einstein condensate of impurity atoms can be observed, as well as waves of matter.

In addition, in superfluid helium there is a possibility of the existence of so-called zero vacancies similar to those in solid helium [7]. We take into account that the wave function of $^4$He atoms in superfluid $^4$He consists on two parts: 1) a zero momentum ($\vec{k}=0$) part of the Bose-Einstein condensate and 2) a momentum-dependent part. The first, condensate part has a long-range order; the second part do not posses the long-range order but it has short-range order characterized by a sharp peak of the pair correlation function of $^4$He atoms at the distance 3.6 Å and the coordination numbe close to $z=12$. The density of the condensate in the superfluid $^4$He even at $T \ll T_C = 2.17\,\text{K}$ is a small fraction of the density of the liquid: only 7% of the atoms are in a long-range ordered condensate state with $\vec{k}=0$ [18]. On the contrary, the short-range fraction includes 93% of all $^4$He atoms and represents the main part of the superfluid helium. This allows us to assume, by analogy with solid $^4$He, that there may be a superfluid $^4$He state with a slightly reduced amplitude of its short-range ordered part. The last reduction corresponds to the presence of zero vacancies.

Zero vacancies must be stable at low temperatures, since removal by "evaporation" from the surface requires their localization in at least one dimension, which requires energy. The latter can come from the phonons of the normal component. At a temperature $T < 1\,\text{K}$, the normal component makes up less than 0.5 percent of the total liquid. In contrast, the entire volume of solid $^4$He contains phonons. Significantly fewer phonons and increased localization energy should make zero vacancies in superfluid $^4$He more stable than in solid $^4$He.

It should be noted that, in our opinion, the existence of zero vacancies in the superfluid $^4$He is indicated by a strong deviation of the shape of rotating large droplets of the superfluid $^4$He from the spherical one observed experimentally in [19-21] (see also [22]): in these works not only oblate spheroidal, but also prolate triaxial shape of raplidly rotating droplets were oserved. Such strong deviations in the shape of a rotating drop from a spherical one can be explained by the accumulation of zero vacancies on the perifery of the drop, where the rotation speed is greatest - such accumulation allows minimizing the energy of the rotating droplet.

**Acknowledgements.** This work was supported by the Estonian Research Council project PRG347.